\documentclass[aps,preprint,prd,showpacs,nofootinbib]{revtex4}
\pdfoutput=1

\usepackage{amsmath,amssymb}
\usepackage{graphicx,subfigure}
\usepackage{color,textcomp}
\usepackage[colorlinks,linkcolor=blue,anchorcolor=blue,citecolor=blue]{hyperref}
\usepackage{ulem} 

\def\lf{\left}
\def\rt{\right}

\def\blue{\color{blue}}

\def\be{\begin{equation}}
\def\ee{\end{equation}}
\def\ba{\begin{eqnarray}}
\def\ea{\end{eqnarray}}

\begin{document}

\title{Trans-Planckian censorship of multistage inflation and dark energy }
\author{Hao-Hao Li$^{1}$\footnote{\texttt{\blue lihaohao18@mails.ucas.ac.cn}}}
\author{Gen Ye$^{1}$\footnote{\texttt{\blue yegen14@mails.ucas.ac.cn}}}
\author{Yong Cai$^{1}$\footnote{\texttt{\blue caiyong13@mails.ucas.ac.cn}}}
\author{Yun-Song Piao$^{1,2}$\footnote{\texttt{\blue yspiao@ucas.ac.cn}}}
\affiliation{$^1$ School of Physics, University of Chinese Academy of
    Sciences, Beijing 100049, China}

\affiliation{$^2$ Institute of Theoretical Physics,
    Chinese Academy of Sciences, P.O. Box 2735, Beijing 100190, China}

\begin{abstract}

We explore the bound of the trans-Planckian censorship conjecture on
an inflation model with multiple stages. We show that if the
first inflationary stage is responsible for the primordial
perturbations in the cosmic microwave background window, the $e$-folding number of each
subsequent stage will be bounded by the energy scale of the first
stage. This seems to imply that the lifetime of the current era of accelerated expansion (regarded as one of the multiple
inflationary stages) might be a probe for distinguishing inflation
from its alternatives. We also present a multistage inflation
model in a landscape consisting of anti-de Sitter vacua separated
by potential barriers.

\end{abstract}

\maketitle
\tableofcontents

\section{Introduction}

Inflation
\cite{Guth:1980zm,Starobinsky:1980te,Linde:1981mu,Albrecht:1982wi},
which sets the initial conditions of hot big bang (BB)
cosmology, is a popular paradigm for the early Universe. The
evolution of the Universe must be implemented in a UV-complete
effective field theory (EFT). It was argued in Refs.
\cite{Ooguri:2006in,Obied:2018sgi} that such EFTs should satisfy
the swampland conjectures. Also importantly, it was pointed out in Refs.
\cite{Brandenberger:2000wr,Martin:2000xs} that the length scales
of fluctuations we observe today might be smaller than the Planck length
in the inflationary phase if inflation lasts long enough, which is the so-called
``trans-Planckian" problem.

Recently, a new swampland condition, i.e., the trans-Planckian censorship
conjecture (TCC), was proposed in Ref. \cite{Bedroya:2019snp}, which
states that the sub-Planckian fluctuations should never cross
their Hubble scale to become classical; otherwise, the
corresponding EFT will belong to the swampland. This actually
suggests that the ``trans-Planckian" problem never happened in a
UV-complete EFT. In other words, a cosmological model that suffers
such a problem is in the swampland. Requiring that the length scale of the sub-Planckian
perturbation will never be larger than the
Hubble scale is equivalent to
    \begin{gather}\label{TCC1}
        \frac{a(t)}{a^{\text{ini}}}l_{\text{P}} < \frac{1}{H(t)}
    \end{gather}
in an expanding Universe,
where $l_{\text{P}}$ is the Planck length. The immediate implications of the TCC for inflation, the early Universe and other aspects in cosmology have
been studied in Refs.
\cite{Bedroya:2019tba,Cai:2019hge,Tenkanen:2019wsd,Das:2019hto,Mizuno:2019bxy,Brahma:2019unn,Dhuria:2019oyf,Torabian:2019zms,Cai:2019igo,Schmitz:2019uti,Kadota:2019dol,Berera:2019zdd,Brahma:2019vpl,Okada:2019yne,Laliberte:2019sqc,Goswami:2019ehb,Lin:2019pmj}.

According to the TCC (\ref{TCC1}), inflation can only last for a limited
$e$-folding number \be \int_{t^{\text{ini}}}^{t^{\text{end}}}H dt<\ln{M_{\text{P}}\over
H^{\text{end}}}\,,\label{eq:tcc01} \ee where the superscripts ``ini" and
``end" represent the beginning and ending of this stage
(i.e., inflation), respectively. Thus, if the TCC is correct, it is not
sufficiently reasonable to set the initial state of the
perturbation modes as the Bunch-Davis state, which is required to explain the
observations in the cosmic microwave background (CMB) window. However, as shown in
Ref. \cite{Cai:2019hge}, a past-complete pre-inflation era can
automatically prepare such initial states.

Furthermore,
the TCC indicates that the energy scale of slow-roll inflation will reduce to $
H_{\text{inf}}\sim 0.1$ GeV, which results in a tensor-to-scalar ratio $
r< 10^{-30} $ in the CMB window, provided  the radiation prevailed
immediately after inflation \cite{Bedroya:2019tba}. However, a
nonstandard post-inflationary history\textemdash in particular, that of a
multistage inflation scenario\textemdash will alleviate this TCC constraint
\cite{Mizuno:2019bxy,Dhuria:2019oyf,Torabian:2019zms,Berera:2019zdd} and might
push $r$ up to $r\lesssim 10^{-8}$.
Therefore, it is interesting to investigate the cosmological scenarios
with multistage inflation, see also earlier studies in, e.g., Refs.
\cite{Dvali:2003vv,Burgess:2005sb,Liu:2009pk}.\footnote{Multistream inflation
\cite{Li:2009sp,Li:2009me,Wang:2010rs,Afshordi:2010wn} might also
be interesting. } The corresponding EFTs should satisfy not only
the TCC, but also other swampland conjectures
\cite{Ooguri:2006in,Obied:2018sgi}.

In this paper, we explore the
bound of the TCC on the multistage inflation scenario and build a
multistage inflation model in a landscape consisting of anti-de
Sitter (AdS) vacua. Also, if the TCC is correct, it seems that the
lifetime of the current accelerated expansion (as one stage of
a multistage inflation scenario) might be shorter than expected in
Ref. \cite{Bedroya:2019snp}.

\section{multistage Inflation}



\subsection{Multistage inflation with the TCC}\label{sec:TCC-inf}

\begin{figure}[tbp]
    \includegraphics[width=4in]{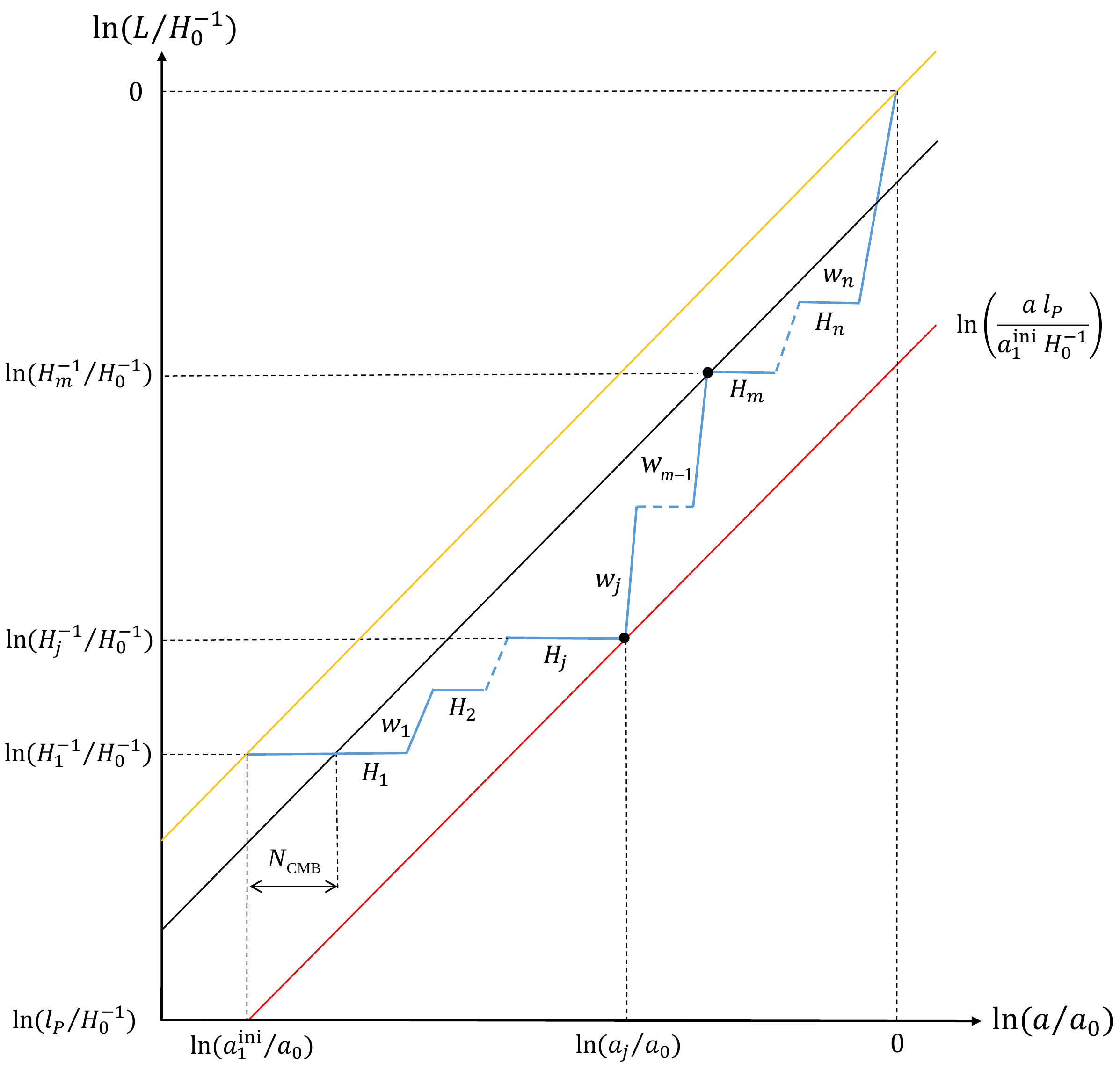}
\caption{The sketch of multistage inflation confronted with the
TCC. Here, $L\propto a$ denotes the physical length and $ H_0 $ is
the Hubble parameter at present. The black line with slope equal to $1$
represents the evolution of the perturbation mode with the largest
wavelength that can be observed at present. The red line
represents the evolution of the perturbation mode with a
physical wavelength equal to the Planck length at the beginning of
the first inflationary stage. The blue polyline represents the evolution
of the Hubble radius, where we have assumed that the $ H_i $'s are almost
constant during each stage of inflation and $w_i>-1/3$.
    }
    \label{horizo}
\end{figure}

Considering a multistage inflation scenario preceding the hot BB expansion, we set the Hubble parameter during the $i$th stage of inflation as a constant $H_i$. We will
assume that after the $i$th stage of inflation, the decelerated
expanding phase has the equation of state $w_i=p_i/\rho_i$ (see
Fig. \ref{horizo} for a sketch), and hence we have $a\sim
t^{\frac{2}{3(1+w_i)}}$. Thus, the slope of $\ln(H^{-1}/H_0^{-1})$
(depicted by the blue polyline) with respect to $\ln(a/a_0)$ at this stage
is $3(1+w_i)/2$, where $H_0$ is the current Hubble constant.
During the decelerated expansion with $w_i
> -1/3$, the Hubble radius will increase faster than the physical
length scale.

The logarithmic interval in comoving wave number of the perturbation modes that cross the horizon during a stage can be defined as
\be
\tilde{N}=\Delta\ln k=\ln\lf({a^{\text{end}}H^{\text{end}}\over a^{\text{ini}}H^{\text{ini}}} \rt)\,, \ee
where $k = aH$ is the comoving scale of the horizon size at time $t$.
We also set the $e$-folding number $N=\ln\lf({a^{\text{end}}\over a^{\text{ini}}}\rt)$, and hence we have
$\tilde{N}=N$ for $H^{\text{end}}=H^{\text{ini}}$. We use $\tilde{N}_i$ (or
$ N_i $) for the $i$th stage of
inflation and $\tilde{N}_{w_i}$ (or $ N_{w_i} $) for the $i$th stage of decelerated expansion with $ w_i>-1/3 $.
Note that $\tilde{N}_i>0$ while $\tilde{N}_{w_i}<0$.


From Fig. \ref{horizo}, we have the following constraints on
$\ln(H^{-1}/H_0^{-1})$.

{\bf Constraint 1:} Perturbation modes with a wavelength
equal to the Planck length at the beginning of the first stage of
inflation will never cross the Hubble horizon. As a result, the
red line [i.e., $\ln\lf({a l_{\text{P}} \over a_{1}^{\text{ini}}H_0^{-1}}\rt)$] in Fig.
\ref{horizo} puts a lower bound on $\ln(H^{-1}/H_0^{-1})$, i.e.,
\be {1\over aH}>{l_{\text{P}}\over a^{\text{min}}}\,,\label{constraint1} \ee where
"min" represents the minimum throughout the history of the
Universe.

{\bf Constraint 2:} Assuming that the first stage of inflation is
responsible for the primordial perturbations in the CMB window,
which should not be polluted by subsequent stages, we will have
an upper bound (the black line in Fig. \ref{horizo}) on
$\ln(H^{-1}/H_0^{-1})$,
i.e., \be {1\over aH}<{e^{-\tilde{N}_{\text{CMB}}}\over
a_{1}^{\text{ini}} H_{1}^{\text{ini}}}\quad \text{for}\quad
a_{1}^{\text{end}}<a<a_{\text{BB}}^{\text{ini}}\,,\label{constraint2} \ee where
$\tilde{N}_{\text{CMB}}$ is the $e$-folding number of the CMB window,
and $a_{\text{inf}}^{\text{end}}$ and $a_{\text{BB}}^{\text{ini}}$ are evaluated at the end of the
first stage of inflation and the beginning of the hot BB expansion,
respectively.

Considering multistage inflation with $n>1$ stages, for the
$j$th stage, we can write down the bound (\ref{constraint1}) imposed
by the TCC as \be
\tilde{N}_j+\sum_{i=1}^{j-1}(\tilde{N}_i+\tilde{N}_{w_i})
=\ln\lf({a_j^{\text{end}}H_j^{\text{end}}\over a_{1}^{\text{ini}}H_{1}^{\text{ini}}}
\rt)<\ln\lf({M_{\text{P}}\over H_{1}^{\text{ini}}}\rt)\,, \label{tildeN}\ee where
we have used $\ln(a_j/a_0)-\ln(a_{1}^{\text{ini}}/a_0) <
\ln(H_j^{-1}/H_0^{-1})-\ln(l_{\text{P}}/H_0^{-1})$ (note that slope of the
red line in Fig. \ref{horizo} is $1$) and $H_{1}^{\text{ini}} \simeq
H_{1}^{\text{end}}$; see also Ref. \cite{Berera:2019zdd}. We have set the Planck mass $M_{\text{P}} = l_{\text{P}}^{-1}$. Thus,
for $n=1$, i.e., single-stage inflation, we have
$\tilde{N}_1<\ln\lf({M_{\text{P}}\over H_{1}^{\text{ini}}}\rt)$ or equivalently $N_1<\ln \lf({M_{\text{P}}\over H_{1}^{\text{end}}}\rt)$, which is
consistent with the result in
Refs. \cite{Bedroya:2019snp,Bedroya:2019tba}.

Based on Eqs.
(\ref{constraint2}) and (\ref{tildeN}), for multistage
inflation, the $j$th stage of inflation
($j\neq 1$) can be bounded as \be \tilde{N}_j<\ln\lf({M_{\text{P}}\over
H_{1}^{\text{ini}}}\rt)-\tilde{N}_{\text{CMB}}\,\label{con:Nj} \ee for
$a_{1}^{\text{end}}<a<a_0$. Therefore, the lifetime of the $j$th stage
inflation is strictly bounded by the energy scale of the first stage of
inflation and the width of the CMB window. Similarly, for a
decelerated expanding stage, e.g., the stage marked by $w_{j-1}$,
we have \be
-\tilde{N}_{w_{j-1}}={1\over2}(1+3w_{j-1})N_{w_{j-1}}<\ln\lf({M_{\text{P}}\over
H_{1}^{\text{ini}}}\rt)-\tilde{N}_{\text{CMB}}\, \label{Nwj}\ee for
$a_{1}^{\text{end}}<a<a_0$. Note that $\tilde{N}_{w_{j-1}}<0$.





Therefore, if the TCC is correct and the first stage of inflation is
responsible for the primordial perturbations in the CMB window, the
$e$-folding number that the $j$th stage ($j\neq 1$, accelerated or decelerated) can last is bounded by the energy scale
(or $H_{1}$) of the first stage of inflation; see also Ref. \cite{Berera:2019zdd}. 

It is also possible that the perturbation modes of the present horizon scale $k_0=a_0H_0$ exit the horizon a few $e$-folds later than the beginning of the first stage of inflation. Furthermore, the CMB scale perturbation modes could even exit the horizon during the $i$th stage of inflation, where $i > 1$, as long as these modes are deep inside the horizon before the $i$th stage \cite{Cai:2019hge}. In these situations, the bounds (\ref{constraint1}) and  (\ref{tildeN}) still hold. However, since $a_{1}^{\text{ini}} H_{1}^{\text{ini}}$ is no longer the comoving wave number of the largest CMB scale perturbation modes, i.e., $k_0\neq a_{1}^{\text{ini}} H_{1}^{\text{ini}}$, the constraint (\ref{constraint2}) should be modified by replacing $a_{1}^{\text{ini}} H_{1}^{\text{ini}}$ with $k_0$. 
As a result, the bounds on $\tilde{N}_j$ and $\tilde{N}_{w_{j-1}}$ for $a_{i}^{\text{end}}<a<a_0$ will be tighter, since the right-hand sides of the bounds (\ref{con:Nj}) and (\ref{Nwj}) should be corrected by subtracting the term $\ln\lf({a_0H_0\over a_{1}^{\text{ini}} H_{1}^{\text{ini}}}\rt)$. Hence, the inequalities (\ref{con:Nj}) and (\ref{Nwj}) are still valid.

\subsection{Multistage inflation in the AdS landscape}

The landscape consists of all EFTs with a consistent
UV completion, which might be from various compactifications of
string theory. Otherwise, the set of corresponding EFTs is
called the swampland. The
distance conjecture $|\Delta\phi|/M_{\text{P}}<\Delta \sim {\cal O}(1)$
\cite{Ooguri:2006in} and the dS conjecture
$M_{\text{P}}|\nabla_{\phi}V|/V>c\sim {\cal O}(1)$ \cite{Obied:2018sgi} [or
the refined dS conjecture  $M_\text{P}^2\frac{\text{min}\left(
        \nabla_{\phi}\nabla_{\phi} V \right)}{V} \leq -c'\sim {\cal O}(1)$]
\cite{Ooguri:2018wrx,Garg:2018reu} have been proposed as the swampland criteria for EFTs. Recently,
many efforts have been made to confront the inflation scenario with the
swampland conjectures; see, e.g., Refs. 
\cite{Achucarro:2018vey,Kehagias:2018uem,Matsui:2018bsy,Kinney:2018nny,Brahma:2018hrd,Motaharfar:2018zyb,Ashoorioon:2018sqb,Das:2018rpg,Kinney:2018kew}.

\begin{figure}[tbp]
\subfigure[ $ V(\Phi) $ ]
{\includegraphics[scale=2,width=0.45\textwidth]{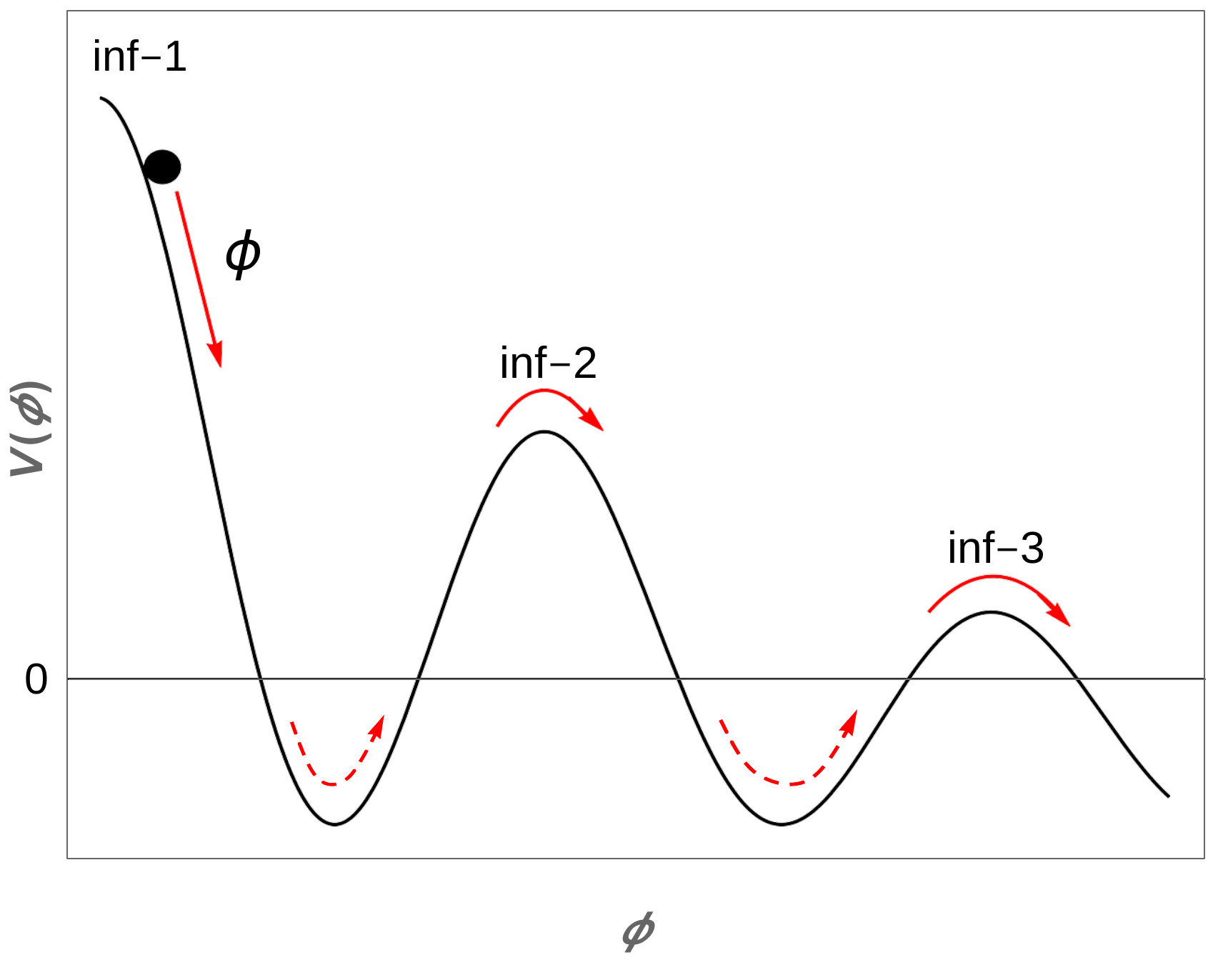}} \quad
\subfigure[ $ H(t) $ ]
{\includegraphics[scale=2,width=0.45\textwidth]{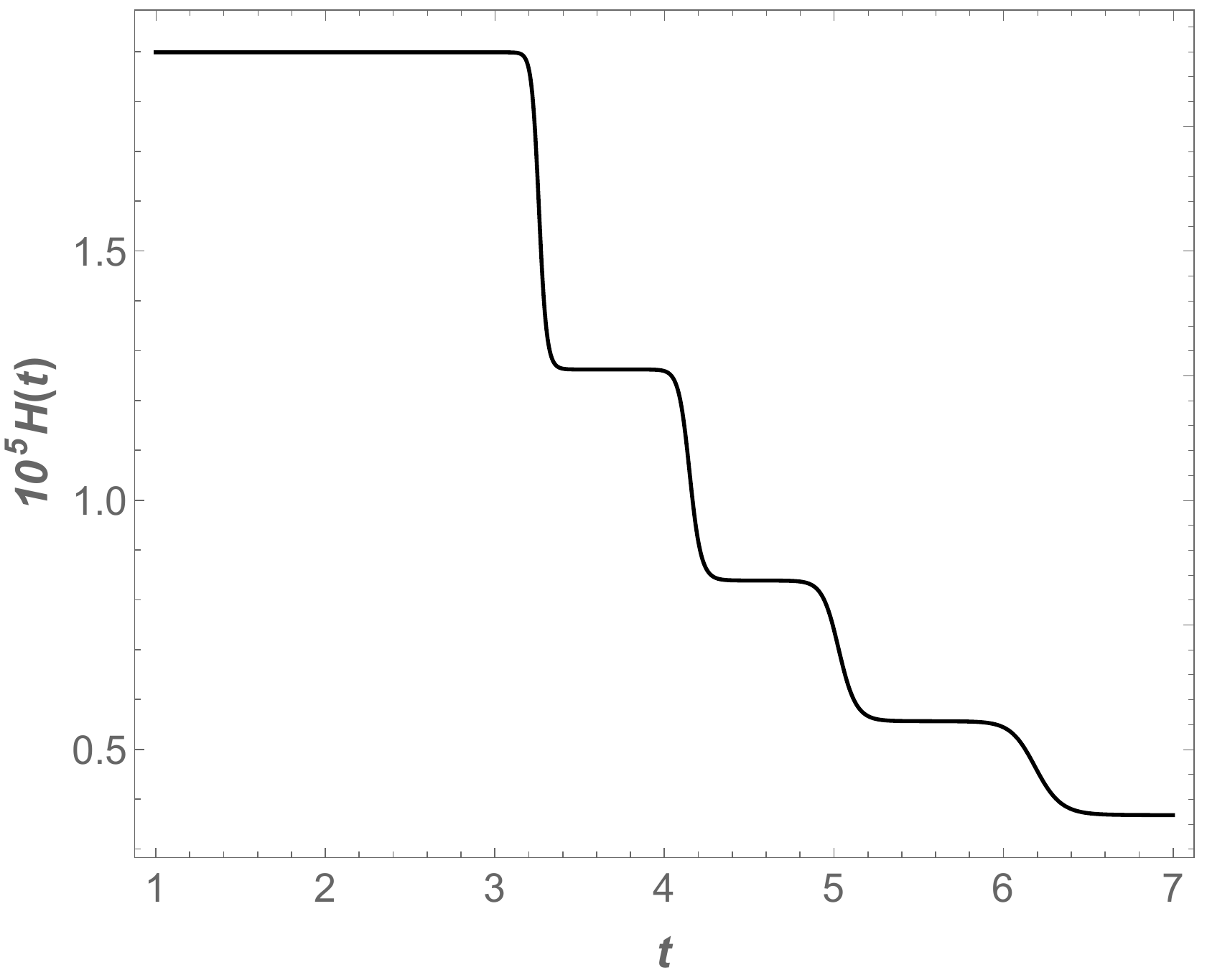}}
\caption{(a) Sketch of the AdS landscape. (b) The
evolution of the Hubble parameter with the potential
(\ref{potential}), in which  $\Lambda = 10^{-12} M_\text{P}^4 $, $V_0=8*10^{-10}M_\text{P}^4$, $\beta=0.642M_\text{P}$ and
    $M=12\sqrt{V_0}M_\text{P}^{-1}$. These parameters satisfy the
restrictions \eqref{DSC} and \eqref{dSSC} of the swampland
conjectures. } \label{background}
\end{figure}

The multistage inflation model (in which a high-scale inflation
is followed by many low-scale inflations) might be interesting,
since it helps to alleviate the bound of the TCC on the energy scale
of inflation in the CMB window. Provided that the string landscape
mostly consists of AdS vacua, we might have such a multistage
inflation model; see Fig. \ref{background}(a). Each stage of
slow-roll inflation only happened around the maxima of the
potential barriers separated
by AdS vacua ($\Delta \phi<M_{\text{P}}$ for each $V>0$ region).\footnote{See also hilltop inflation
		\cite{Senoguz:2004ky,Boubekeur:2005zm,Kohri:2007gq}.}
Between the $j$th and $(j+1)$th stages, the field will rapidly
roll over an AdS vacuum.

As an illustrative example, we model the effective potential
$V_{\text{eff}}(\phi)$ as
\begin{gather}
V_{\text{eff}}(\phi)=V_{0}\left( 1-\cos\left( \frac{M}{\sqrt{V_{0}}}\phi
\right) \right)e^{-\phi/\beta} - \Lambda,\label{potential}
\end{gather}
which is similar to Fig. \ref{background}(a) (see also Ref. [48]), where $\Lambda =$ const $>0$
forces minima of the
potential to be AdS, and $V_{0}$, $M$ and $ \beta $ are
positive constants that set the amplitude, period and
envelope curve of $V_{\text{eff}}(\phi)$, respectively. The AdS potential
(\ref{potential}) must satisfy the swampland distance conjecture
as well as the de Sitter (or refined dS) conjecture. Considering that $
1-\cos \left( \frac{M}{\sqrt{V_0}}\phi \right) = 0 $ and the AdS
well has a width, we have $ \Delta \phi < \frac{2\pi\sqrt{V_0}}{M}
$ for $ V(\phi)>0$. Hence, the swampland distance conjecture
$|\Delta\phi|/M_{\text{P}}< {\cal O}(1)$ requires
\begin{gather}
\frac{\sqrt{V_0}}{M} < {\cal O}(1)\frac{M_{\text{P}}}{2\pi}\,. \label{DSC}
\end{gather}
Confronting the maxima of the potential (\ref{potential}) [at which $\cos\left( \frac{M}{\sqrt{V_0}}\phi\right) \approx -1 $] with
the refined de Sitter conjecture  $M_\text{P}^2\frac{\text{min}\left(
        \nabla_{\phi}\nabla_{\phi} V \right)}{V} \leq -c'\sim {\cal O}(1)$,
we have
\begin{gather}
2V_0/\beta^2-M^2 < -{\cal O}(1)\lf({{2V_0-\Lambda}\over
{M_{\text{P}}^2}}\rt) \label{dSSC}\,.
\end{gather}

We will show the possibility of multistage inflation in
an AdS landscape like Eq. (\ref{potential}). For simplicity, we assume that
each stage of slow-roll inflation happens around one of the maxima
of the potential (which is much larger than the depth of the AdS well, i.e., $V^j_{\text{inf}}\gg
|\Lambda|$) and ends when the inflaton rolls rapidlly down the hill.
When the $j$th stage of inflation ends, the potential energy
$V_{\text{inf}}^j$ will be thoroughly converted to ${\dot \phi}^2$, so
that the field can rapidly roll over an AdS well. During this era,
which corresponds to $w_{j}\simeq 1$, we have $\dot{\phi}\sim
a^{-3}$ and $\rho_{\phi}\approx \dot{\phi}^2/2\sim a^{-6}$.
According to $3M_{\text{P}}^2H^2=\rho_{\phi}$, we get $\dot{\phi}\approx
\sqrt{2/3}{M_{\text{P}}/t}$. Note that this result is valid only if
${\dot\phi}^2(t)\gg |\Lambda|$. After it rapidly rolls over the
AdS well, the field must be able to slowly climb over the next maximum of the
potential (${\dot \phi}^2\ll V_{\text{inf}}^{j+1}$), so that the
$(j+1)$th stage of inflation can happen. During the period ${\dot
\phi}^2\gg |V_{\text{eff}}|$, the shift of $\phi$ is approximately \be
\Delta\phi=\int_{t^{\text{ini}}}^{t^{\text{end}}}\dot{\phi}dt\approx
\sqrt{2\over3}M_{\text{P}}\ln\lf({t^{\text{end}}\over t^{\text{ini}}}
\rt)\,.\label{eq:deltaphi} \ee 

Based on the above analysis, 
the ratio between the potential energy of the $(j+1)$th and
$j$th stages of inflation is ${V_{\text{inf}}^{j+1}/ V_{\text{inf}}^{j}}\approx
\lf({\dot{\phi}^{\text{end}}/ \dot{\phi}^{\text{ini}}}\rt)^2\approx \lf({t^{\text{ini}}/ t^{\text{end}}}
\rt)^2$. Using Eq. (\ref{eq:deltaphi}), we find \be
{V_{\text{inf}}^{j+1}\over V_{\text{inf}}^{j}}\approx \exp\lf(-\sqrt{6}
{\Delta\phi\over M_{\text{P}}} \rt)\,.\label{eq:VV} \ee Therefore, the AdS
landscape (\ref{potential}) is viable for multistage
inflation. Actually, from (\ref{potential}) and (\ref{eq:VV}) we
find $\beta\approx {\cal O}(M_{\text{P}}/\sqrt{6})\approx {\cal
O}(0.4M_{\text{P}})$. 
However, it can be inferred that the value of $\beta$ is fine-tuned so that the inflaton does not overshoot the potential maxima at a large velocity or become trapped in the AdS vacuum. The numerical calculation hints that there exists an amount of parameter fine-tuning at least of order $10^{-2}M_{\text{P}}$ for $\beta$ to prevent both things from happening. How to alleviate the fine-tuning in a realistic multistage inflation model requires further investigation.

Here, as an example, we plot the evolution of the background (the Hubble parameter) with
the AdS potential (\ref{potential}) in Fig. \ref{background}(b),
which is clearly multistage inflation. The parameters $V_{0}$,
$M$, $ \beta $, and $\Lambda$ used in Fig. \ref{background}(b)
satisfy the bounds \eqref{DSC} and \eqref{dSSC}. Here we have
assumed $V^i_{\text{inf}}\gg |\Lambda|$, and thus the ratio
${H_{\text{inf}}^{j+1}/H_{\text{inf}}^{j}}$ is small. Generally, the AdS minima
have different values. Consequently, we may not have $V^i_{\text{inf}}\gg |\Lambda|$
for each $V^i_{\text{inf}}$, which will result in a larger
${H_{\text{inf}}^{j+1}/H_{\text{inf}}^{j}}$.

\section{Implications for Dark Energy}

\begin{figure}[tbp]
    \includegraphics[width=4in]{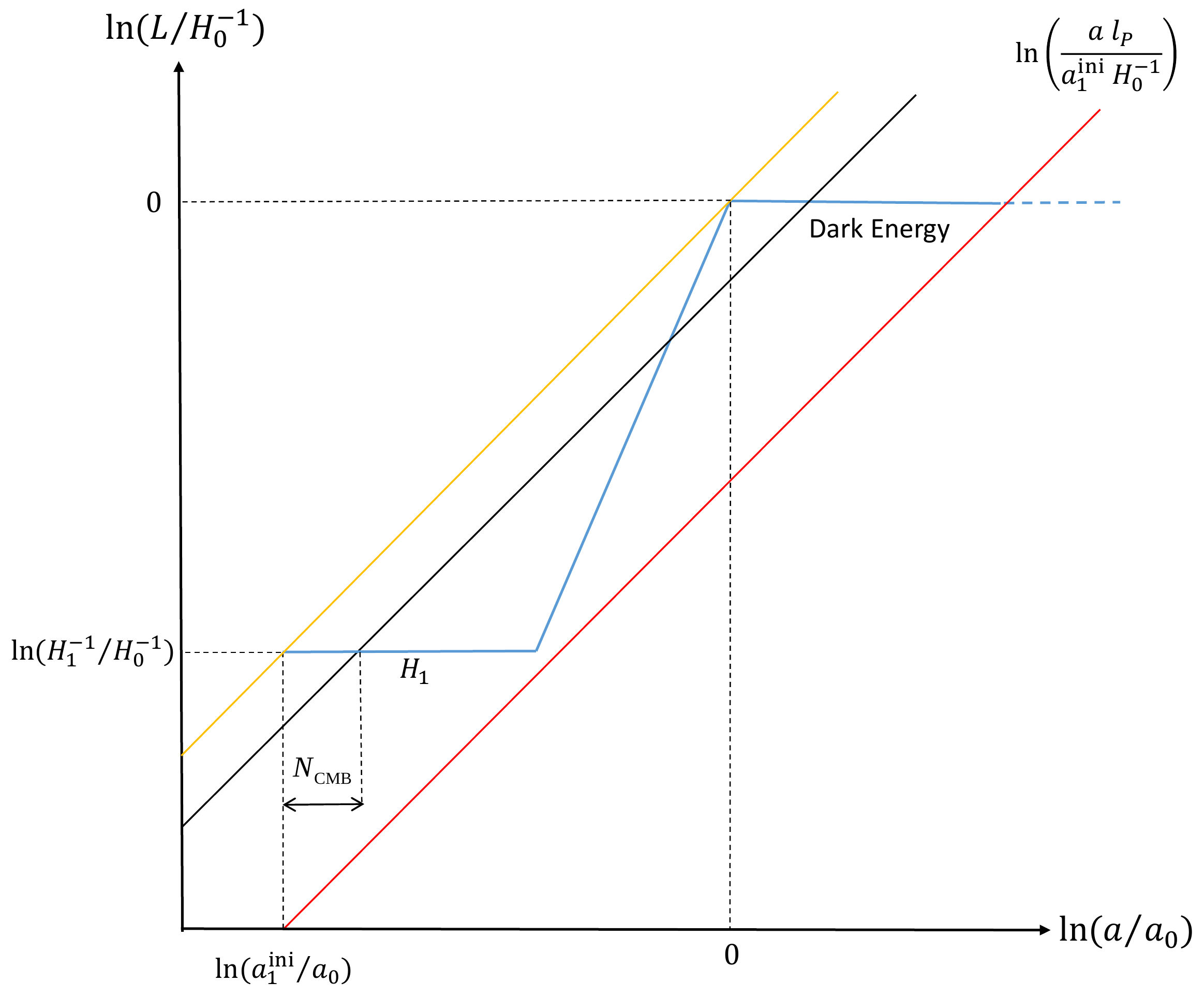}
    \caption{Sketch of the dark energy era confronted with the TCC.} \label{DE}
\end{figure}

The Universe is currently in an accelerated phase of expansion. Nearly $ 70 $\% of the Universe is made up of dark energy (DE)
\cite{Aghanim:2018eyx} and its equation-of-state parameter is
approximately $ w \simeq -1 $. Thus, this accelerated stage of expansion has similar behavior as primordial inflation, which
hence may be thought of as one stage of a multistage inflation scenario,
see Fig. \ref{DE}.
According to Sec. \ref{sec:TCC-inf}, the TCC will put a bound on the lifetime of the DE era.

For simplicity, we assume that after the first stage of inflation
($H_1=H_{\text{inf}}$) the Universe experiences the hot BB evolution
up to the present DE era (the second stage of inflation, $H_2=H_{\text{DE}}$). Thus
setting $j=2$ in Eq. (\ref{tildeN}), the TCC bound on the DE era is
\be \tilde{N}_{\text{DE}}< \ln\lf({M_{\text{P}}\over
H_{\text{inf}}^{\text{ini}}}\rt)\,,\label{con:DE} \ee since
$\tilde{N}_{w_1}=-\tilde{N}_1$. Here we do not need to make
allowance for the CMB window, which actually corresponds to setting
$a_{\text{DE}}^{\text{ini}}H_{\text{DE}}^{\text{ini}}\approx a_{\text{inf}}^{\text{ini}}H_{\text{inf}}^{\text{ini}}$. 
Therefore,
if $H_{\text{DE}}=H_0$ is approximately constant, the lifetime  of the current DE era is \be \Delta t<{1\over H_0}\ln\lf({M_{\text{P}}\over
H_{\text{inf}}^{\text{end}}}\rt), \label{con:DE1}\ee 
which is far smaller than the
expected value ${1\over H_0}\ln({M_{\text{P}}\over H_{0}})$, since
$H_{\text{inf}}^{\text{end}}\gg H_0$ (even if $H_{\text{inf}}^{\text{end}}$ is reduced to
$\sim 0.1 $ GeV \cite{Bedroya:2019tba}). Intriguingly, if the TCC
is correct, it seems that the fate of our Universe is fixed by the properties of primordial inflation, rather than DE itself.

It should be mentioned that we have assumed that the present horizon scale $k_0$ is the same as the horizon size at the beginning of inflation, i.e., $k_0=a_0H_0=a_{\text{inf}}^{\text{ini}}H_{\text{inf}}^{\text{ini}}$. If the perturbation mode with comoving wave number $k_0$ exits the horizon a few $e$-folds after the beginning of inflation, then the bounds on $\tilde{N}_{\text{DE}}$ and $\Delta t$ will be tighter, since the right-hand sides of Eqs. (\ref{con:DE}) and (\ref{con:DE1}) should be subtracted by terms $\ln\lf({a_{\text{DE}}^{\text{ini}}H_{\text{DE}}^{\text{ini}}\over a_{\text{inf}}^{\text{ini}}H_{\text{inf}}^{\text{ini}}} \rt)$ and ${1\over H_0}\ln\lf({a_{\text{DE}}^{\text{ini}}H_{\text{DE}}^{\text{ini}}\over a_{\text{inf}}^{\text{ini}}H_{\text{inf}}^{\text{ini}}} \rt)$, respectively. Therefore, the bounds (\ref{con:DE}) and (\ref{con:DE1}) are still valid in this situation, but the actual allowed lifetime of the DE era should be shorter for an amount of ${1\over H_0}\ln\lf({a_{\text{DE}}^{\text{ini}}H_{\text{DE}}^{\text{ini}}\over a_{\text{inf}}^{\text{ini}}H_{\text{inf}}^{\text{ini}}} \rt)$ than that estimated from Eq. (\ref{con:DE1}).


It is also possible that alternative models of inflation could be the origin of primordial perturbations that are consistent with observations (e.g., Refs. \cite{Khoury:2001wf,Finelli:2001sr,Brandenberger:1988aj,Piao:2003ty,Creminelli:2010ba}).
It would be interesting to see what would happen to $\tilde{N}_{\text{DE}}$ with these alternatives. Considering a nonsingular bouncing scenario, we have
$a^{\text{min}}=a_{\text{b}}$, where $a_{\text{b}}$ is evaluated at the bounce point.
According to Eq. (\ref{constraint1}), assuming that the Universe will
start to hot-BB-like expand immediately after the bounce (i.e.,
$a_{\text{b}}\approx a_{\text{BB}}^{\text{ini}}$), we have \be
\tilde{N}_{\text{DE}}<\ln\lf({a_{\text{b}}/l_{\text{P}}\over a_{\text{DE}}^{\text{ini}}H_{\text{DE}}^{\text{ini}}}\rt)
\approx \lf|\tilde{N}_{\text{BB}}\rt|+\ln\lf({M_{\text{P}}\over
H_{\text{BB}}^{\text{ini}}}\rt)\,,\label{NDE-bounce} \ee where
$\tilde{N}_{\text{BB}}=\ln\lf({a_{\text{DE}}^{\text{ini}}H_{\text{DE}}^{\text{ini}}\over
a_{\text{BB}}^{\text{ini}}H_{\text{BB}}^{\text{ini}} } \rt)<0$ is the logarithmic interval in comoving wave number of the perturbation modes entering the horizon during the hot
BB evolution before the DE era. Here, Eq. (\ref{NDE-bounce}) is
applicable whether we assume an ekpyrosis
\cite{Khoury:2001wf} or matter bounce \cite{Finelli:2001sr}  scenario. Also,
for the slow-expansion scenario (in which the expansion is ultra-slow with
$\epsilon=-{\dot H}/H^2\ll -1$ before the hot BB evolution)
\cite{Piao:2003ty,Creminelli:2010ba} the result is similar.


Therefore, if the
Universe rapidly reheated after the end of inflation or its alternatives,  such that $H^{\text{end}}_{\text{inf}}=H_{\text{BB}}^{\text{ini}}$,
and also experienced the same hot BB expansion, the DE era
in the inflation scenario would be much shorter than that in
alternatives to inflation. This is because the TCC bound on inflation is much tighter than that on its
alternatives, as pointed out in
Ref. \cite{Bedroya:2019tba}. Interestingly, if the TCC is correct,
the lifetime of the DE era might be a probe for distinguishing different possibilities for the origins of the Universe.


\section{Discussion}

As one possible criteria for consistent EFTs, the TCC sets an
upper bound on the energy scale of inflation, which renders $r$ in the CMB window negligibly small. Recently, the implications of the TCC for inflation have inspired studies of multistage inflation.

We showed that if the TCC is correct and the first stage of inflation
is responsible for the primordial perturbations in the CMB window, the $e$-folding number of the $j$th stage ($j\neq 1$) will be
bounded by the energy scale of the first stage of inflation.
Regarding the DE era as one stage of a multistage inflation
scenario, we pointed out that in the inflation scenario the lifetime
of the current DE era is far smaller than expected, which might be a
probe for distinguishing inflation from its alternatives. 
Our result may also place constraints on EFTs of DE, which would be an interesting topic for future work.

By assuming that the landscape of EFTs consists mostly of AdS vacua,
we presented a multistage inflation model in the AdS landscape (with
AdS vacua separated by barriers $V>0$). Each inflationary stage only happens around the maxima of the potential
barriers. The field will rapidly roll over an
AdS vacuum between two stages. Such a model naturally satisfies all swampland
conjectures. Note that the AdS potential is also present in
past-complete pre-inflation stages \cite{Cai:2019hge}. Thus, if the
TCC is correct, the issues related to the early Universe and DE
(especially in the AdS landscape) would be worthy of further exploration.

\section*{Acknowledgments}

Y. C. is funded by the China Postdoctoral Science Foundation (Grant  No. 2019M650810) and NSFC (Grant No. 11905224).
Y. S. P. is supported by NSFC (Grant Nos. 11575188 and 11690021).

\end{document}